\documentclass[10pt,a4paper]{article}
\usepackage{bm}
\usepackage{outlines}
\usepackage{amsmath}
\usepackage{amsfonts}
\usepackage{amssymb}
\usepackage{mathptmx}
\usepackage{hyperref}
\usepackage{appendix}
\usepackage{dcolumn}
\usepackage{url}
\usepackage{ulem}
\usepackage{cancel}
\usepackage{wasysym}
\usepackage{graphicx,subcaption,caption}
\usepackage{sidecap}
\usepackage{wrapfig}
\usepackage{exercise}
\usepackage{xcolor}
\usepackage{lipsum}
\usepackage{catchfilebetweentags}
\usepackage{lineno}
\usepackage{tikz,pgfplots}\usetikzlibrary{arrows,positioning,shapes,shadows,decorations.markings,decorations.pathmorphing}\tikzset{->-/.style={decoration={markings,mark=at position #1 with {\arrow{>}}},postaction={decorate}}}\tikzset{decay/.style={->,decorate,decoration={snake,amplitude=#1,segment length=1.5mm,post length=#1}}}

\newcommand{\bn}{\begin{align}}
\newcommand{\enl}[1]{\label{#1}\end{align}}

\newcommand{\equ}[2]{\begin{equation}\begin{split}{#1}\end{split}\label{#2}\end{equation}}
\newcommand{\ba}{\begin{equation}}
\newcommand{\bs}{\begin{split}}
\newcommand{\ea}{\end{equation}}
\newcommand{\el}[1]{\label{#1}\end{equation}}
\newcommand{\bsa}{\begin{subequations}}
\newcommand{\esa}{\end{subequations}}
\newcommand{\esl}[1]{\label{#1}\end{subequations}}
\newcommand{\baq}{\begin{eqnarray}}
\newcommand{\eaq}{\end{eqnarray}}

\newcommand{\eq}[1]{Eq.~(\ref{#1})}
\newcommand{\fig}[1]{Fig.~(\ref{#1})}

\newcommand{\bra}{\langle}
\newcommand{\ket}{\rangle}
\newcommand{\td}{\tilde}
\newcommand{\+}{\dagger}

\newcommand{\de}{\delta}

\newcommand{\e}{\eta}
\newcommand{\tx}{\theta}

\newcommand{\q}{\phi}

\usepackage{times,catchfilebetweentags}
\begin{document}
\title{Optical ranging with quantum advantage}
\author{Sankar Davuluri$^{*1}$, Greeshma Gopinath$^{1}$ and Matt J. Woolley$^{2}$\\\textit{$^{1}$\textbf{BITS} Pilani, Hyderabad Campus, 500078, India}\\\textit{$^{2}$School of Engineering and Technology, UNSW Canberra,}\\\textit{Canberra, Australian Capital Territory, Australia.}\\\tt{$^*$sankar@hyderabad.bits-pilani.ac.in}}\date{\today}
\maketitle
\textbf{abstract} The quantum illumination technique requires joint measurement between the idler and the probe reflected from the low-reflective target present in a noisy environment. The joint measurement is only possible with prior knowledge about the target's location. The technique in this article overcomes this limitation by using entanglement and a cross-correlated homodyne measurement. This technique does not require quantum storage of the idler and prior knowledge about the target's distance. The cross-correlation measurement makes this technique completely immune to environmental noise, as the correlation between the idler and the environment is zero. The low reflectivity of the target is negated by increasing the intensity of the reference fields (non-entangled) in the homodyne. Based on heuristic arguments, a lower bound of the target's reflectivity for optimum application of this technique is described.
\section{Introduction}
\par Quantum illumination (QI)~\cite{lloyd-08, genovesa-13, gu-20} has shown that entangled photons can detect the presence of a low reflective target in a noisy background more efficiently than non-entangled photons. The QI protocol is studied using different quantum states like Gaussian states~\cite{shapiro-08,park-22,pirandola-20}, discreet~\cite{zhao-20} and continuous variable entanglement~\cite{assad-21,guo-14}, cat states~\cite{wei-23}, photon number entangled states~\cite{su-yong-lee-22}, $N$-photon entangled states~\cite{kim-21}, and hyper-entanglement~\cite{chandrashekar-21}. There have been several studies~\cite{jeffers-21,erkmen-09,shapiro-17-a} about optimum detection schemes like parametrically amplified idler~\cite{blakely-21} and double homodyne~\cite{su-yong-lee-21}. A remarkable feature of QI is that the residual quantum correlations~\cite{kenchaf-22,park-23} between the probe and the idler will provide a quantum advantage even after losing the entanglement because of the environment~\cite{lloyd-09, shapiro-15, rajgopal-09}. The QI technique is extended to microwave frequencies theoretically~\cite{pirandola-15} and experimentally~\cite{fink-20,balaji-19,huard-23,wilson-19}. Recently, QI has used the quantum eraser concept to make it immune to jamming~\cite{wang-23} by detecting only the locally stored reference fields. The robustness of QI to environment and background noise makes it particularly useful for radar technology. However, QI detection of the low-reflective target in different locations requires appropriately readjusting the idler storage times for each location. If the location of the low-reflective target is unknown, storing the idler exactly until the probe's return is impossible. This limits the application of QI from full-ranging applications like lidar and radar. The lack of quantum ranging function has limited QI to imaging in some studies~\cite{padgett-20}. Nevertheless, QI unquestionably outperforms~\cite{guo-22, chandrashekar-23, helmy-22, helmy-19} the classical detection strategies in noisy environments. Hence, the QI has been modified into several versions, discussed in the next paragraph, to build a quantum radar or lidar.
\par Digitization technique~\cite{fink-20,england-18,balaji-19,wilson-19} circumvents the quantum storage of idler by measuring the signal and probe independently. The measured data is post-processed to complete the QI protocol digitally~\cite{fink-20}. This method can not add ranging functionality because joint measurement of idler and signal photons in the digitized version is required. The authors in Ref.~\cite{balaji-19,wilson-19} studied quantum noise correlated radar. Their experiment introduced quantum noise correlations using entanglement but did not use a low-reflective target for detection and ranging. Instead, their experiment showed entanglement enhanced detection of microwaves in a noisy environment. Later studies~\cite{sussman-19, sussman-20} included a target, but these studies switched to photon coincidence detection, which requires prior knowledge about the target's location. There have been several theoretical proposals~\cite{ren-20, sanz-22, zhuang-21,knott-21} to overcome the range and/or velocity limitation of QI protocol and to study the quantum limits~\cite{kok-21,gu-20,shapiro-17}. Multiary hypothesis testing added ranging functionality to QI by dividing the entire range into multiple blocks~\cite{zhuang-21}; however, an optimum detection scheme for a range with blocks more than two remains an open question. Quantum localization is studied~\cite{ren-20} to add ranging function to QI; however, the $N$ photon entangled probes are highly sensitive to decoherence~\cite{dowling-08} and achieving entangled states with $N>2$ is quite challenging. The quantum-enhanced Doppler lidar~\cite{sanz-22} can provide information about the target's velocity but not distance. 
\par In this paper, we propose combining QI with a cross-correlated homodyne detection scheme using digitization~\cite{fink-20,england-18,balaji-19,wilson-19}. This technique offers the following advantages compared to the standard quantum illumination technique. (i) Quantum storage of the idler field and prior knowledge about the distance of the low-reflective target are not required. (ii) Target detection and ranging efficiency are independent of entangled field intensity and environmental noise. (iii) The low-reflectivity property of the target can be negated by increasing the intensity of the reference fields (non-entangled) in the homodyne. 
\section{Quantum illumination with homodyne cross-correlation}
\begin{figure}[htb]\centering\includegraphics[height=5cm,width=0.7\textwidth]{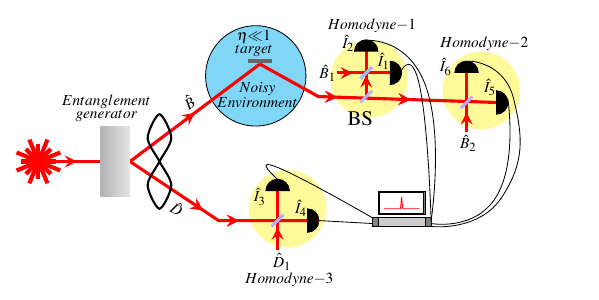}\caption{Quantum illumination with cross-correlated homodyne. $\hat B$ and $\hat D$ are the destruction operators for the entangled fields. $\hat B_1$, $\hat B_2$, and $\hat D_1$ are destruction operators for the reference fields in homodyne setups. $\hat I_i\, (i=1,2...6)$ are intensities at the photo detectors. BS is a beam splitter.}
\label{f101}\end{figure}
Figure~\ref{f101} shows the schematics of the system under study. Destruction operators of each entangled are represented with $\hat B$ and $\hat D$ as shown in \fig{f101}. The $\hat D$ field mixes with a reference field, represented with destruction operator $\hat D_1$, in the homodyne-3 setup shown in \fig{f101}. Then the difference in intensities is given as
\equ{\hat I_4-\hat I_{3}\approx2\sqrt{I_DI_{D1}}\sin\q_1+ [\cos\q_1(\sqrt I\hat Y_D-\sqrt I_{D}\hat Y_{D1})+\sin\q_1(\sqrt I\hat X_D+\sqrt I_{D}\hat X_{D1})],}{423}
where $\q_1$ is the phase difference between $\hat D$ and $\hat D_1$, $\hat Y_O=i(\hat\de_O^\+-\hat\de_O)$, $\hat X_O=\hat\de_O^\++\hat\de_O$ with $\hat\de_O$ as the quantum fluctuation and $\bar O$ as the classical mean of operator $\hat O$ ($O=B,D,D_1)$, $I_D=\bra\hat D^\+\hat D\ket$ and $I=\bra\hat D_1^\+\hat D_1\ket$. The superfix `$\dagger$' represents an adjoint operation. All the optical operators $\hat B$, $\hat D$, and $\hat D_1$ are normalized such that $I$ and $I_D$ represent the number of photons per unit time. By tuning the $\hat D_1$ phase, we set $\q_1=\pi/2$. Then the linearized quantum fluctuation $\hat\tx_D$ in \eq{423} is given as
\equ{\hat \tx_D(t)=[\sqrt{I}\hat X_D(t)+\sqrt{I_{D}}\hat X_{D1}(t)],}{419}
where $t$ is the time. Showing the $t$ dependence explicitly in \eq{419} will be useful later in this draft.
\par On the other hand, $\hat B$ is sent into the environment to probe the presence of a low reflective $\e\ll1$ target. A tiny amount of $\hat B$ will be received if the low reflective target is present; otherwise, only the environmental noise will be received. The received signal is given as $\sqrt\e\hat Be^{i\q_B}+\sqrt{1-\e}\hat E$, where $\hat E$ and $\q_B$ are the destruction operator of the environment and phase acquired by $\hat B$ as it travels through the unknown path in the environment, respectively. This signal is then split into two parts using the beam splitter BS as shown in \fig{f101}. One part mixes with a reference field $\hat B_2e^{i\q_2}$, where $\q_2$ and $\hat B_2$ are the phase and destruction operator of the reference field, in homodyne-2 (see the \fig{f101}). Then the difference in intensities at the photodetectors is given as
\equ{\hat I_{5}-\hat I_{6}=i\sqrt\frac{\e}{2}(\hat B^\+\hat B_2e^{-i\q}-\hat B_2^\+\hat Be^{i\q})+i\sqrt{\frac{1-\e}{2}}(\hat E^\+\hat B_2e^{i\q_2}-\hat B_2^\+\hat Ee^{-i\q_2})\\+\sqrt{\frac{1}{2}}(\hat V^\+\hat B_2e^{i\q_2}+\hat B_2^\+\hat Ve^{-i\q_2}),}{426}
where $\q=\q_B-\q_2$, the superfix `*' represents complex conjugate, and $\hat V$ is the destruction operator for the vacuum field entering through the empty port of BS. By writing $\hat B_2$ as sum of its classical mean $\bar B_2$ and quantum fluctuation $\hat\de_{B_2}$, the linearized fluctuation $\hat\tx_{B2}$ in \eq{426} is given as
\equ{\hat\tx_{B2}(t_1)=&\sqrt{\frac{\e}{2}}\{[\cos\q[\sqrt{I}\hat Y_B(t_1)-\sqrt{I_B}\hat Y_{B2}(t_1)]+\sin\q[\sqrt{I}\hat X_B(t_1)+\sqrt{I_B}\hat X_{B2}(t_1)]\}+\\&i\sqrt{\frac{(1-\e)I}{2}}[\hat E^\+(t_1)e^{i\q_2}-e^{-i\q_2}\hat E(t_1)]+\sqrt{\frac{I}{2}}[\hat V^\+(t_1)e^{i\q_2}+e^{-i\q_2}\hat V(t_1)],}{420}
where $I_B=\bra\hat B^\+\hat B\ket$ and we set $\bra\hat B_2^\+\hat B_2\ket=I$. The reflected field from the BS mixes with another reference field, with destruction operator $\hat B_1$ and phase $\q_2$, in homodyne-1. The $\hat B_1$ is phase-locked with $\hat B_2$, so both have the same phase $\q_2$. The optical path lengths from the BS to $homodyne-1$ and $homodyne-2$ are adjusted to be equal. Then the difference in intensities at the photo-detectors is
\equ{\hat I_{2}-\hat I_{1}=\sqrt\frac{\e}{2}(\hat B^\+\hat B_1e^{-i\q}+\hat B_1^\+\hat Be^{i\q})+\sqrt{\frac{1-\e}{2}}(\hat E^\+\hat B_1e^{i\q_2}+\hat B_1^\+e^{-i\q_2}\hat E)\\+i\sqrt{\frac{1}{2}}(\hat V^\+\hat B_1e^{i\q_2}-\hat B_1^\+e^{-i\q_2}\hat V).}{446}
After writing $\hat B_1$ as sum of its mean $\bar B_1$ and quantum fluctuation $\hat\de_{B_1}$, the linearized fluctuation $\hat\tx_{B1}$ in \eq{446} is 
\equ{\hat\tx_{B1}(t_1)=&\sqrt{\frac{\e}{2}}\{-\sin\q[\sqrt{I}\hat Y_B(t_1)-\sqrt{I_{B}}\hat Y_{B1}(t_1)]+\cos\q[\sqrt{I}\hat X_B(t_1)+\sqrt{I_{B}}\hat X_{B1}(t_1)]\}+\\&\sqrt{\frac{(1-\e)I}{2}}[\hat E^\+(t_1)e^{i\q_2}+e^{-i\q_2}\hat E(t_1)]+i\sqrt{\frac{I}{2}}[\hat V^\+(t_1)e^{i\q_2}-e^{-i\q_2}\hat V(t_1)],}{440}
where $\bar B_1$ is the mean value of $\hat B_1$, and we set $\bra\hat B_1^\+\hat B_1\ket=I$. Note the phase difference is $\q$ in both \eq{440} and \eq{420} as $\hat B_1$ and $\hat B_2$ are phase-locked to have the same phase. Using \eq{423}, \eq{440} and \eq{420}, we can write
\equ{\bra\overline{\hat\tx_D(t)\hat\tx_{B2}(t_1)}\ket=\sqrt{\frac{\e}{2}I^2}[\cos\q\bra\overline{\hat X_D(t)\hat Y_B(t_1)}\ket+\sin\q\bra\overline{\hat X_D(t)\hat X_B(t_1)}\ket],}{430}
\equ{\bra\overline{\hat\tx_D(t)\hat\tx_{B1}(t_1)}\ket=\sqrt{\frac{\e}{2}I^2}[-\sin\q\bra\overline{\hat X_D(t)\hat Y_B(t_1)}\ket+\cos\q\bra\overline{\hat X_D(t)\hat X_B(t_1)}\ket],}{429}
where $\bra\overline{\hat O_1\hat O_2}\ket=\bra\hat O_1\hat O_2+\hat O_2\hat O_1\ket/2$ is the symmetrization operation with $\hat O_1=\hat\tx_D, \hat X_D,\hat Y_D$ and $\hat O_2=\hat\tx_{B_1},\hat\tx_{B_2},\hat Y_B,\hat X_B$. Note that the cross-correlations in \eq{429} and \eq{430} are immune to environmental noise. The \eq{429} and \eq{430} can be non-zero only if $\hat B$ and $\hat D$ are quantum correlated. A non-zero value for \eq{429} and \eq{430} establishes the presence of the low-reflective target. The $\q$ in \eq{429} and \eq{430} is an unknown variable as the exact trajectory of $\hat B$ is unknown. The $\q$ dependence can be eliminated from \eq{429} and \eq{430} by rewriting them as
\equ{\bra\overline{\hat\tx_D(t)\hat\tx_{B1}(t_1)}\ket^2+\bra\overline{\hat\tx_D(t)\hat\tx_{B2}(t_1)}\ket^2=\frac{\e}{2}I^2[\bra\overline{\hat X_D(t)\hat Y_B(t_1)}\ket^2+\bra\overline{\hat X_D(t)\hat X_B(t_1)}\ket^2].}{441}
Measurement of cross-correlations such as in \eq{441} is discussed in Ref.~\cite{girvin-10}. The $\hat X$ and $\hat Y$ quadratures of the signal and idler fields are proportional to the voltage quadratures in the homodyne~\cite{fink-20,wilson-19} measurement. The \eq{441} is evaluated by scanning the pre-recorded homodyne-3 measurement against the homodyne-1 and homodyne-2 measurement records. As $\bra\overline{\hat X_B(t)\hat Y_D(t_1)}\ket\propto\de(t-t_1)$, a non-zero value for \eq{441} implies that the distance of the low-reflective target is $(t-t_1)c/2$. We assumed that the low-reflective object is equidistant from the receiver and sender and $c$ is the velocity of light.
\par Generally, the $\e$ of a non-cooperative target is not known and is very small. The small value of $\e$ in \eq{441} can be compensated by increasing the intensity of the reference fields in the homodynes. It is possible to set $I\ggg1$ so that the product $\e I^2\geq2$ for $\e\ll1$ in \eq{441}. If the correlation in \eq{441} is greater than or equal to $\bra\overline{\hat Y_D(t)\hat Y_B(t_1)}\ket^2+\bra\overline{\hat Y_D(t)\hat X_B(t_1)}\ket^2$, the presence of a target with $\e\geq2/I^2$ can be confirmed and its distance can be measured. Hence, it is safe to claim that the technique described in this work can detect and range targets with $\e\geq2/I^2$.
\subsection{Optimization}The correlations in \eq{441} are dependent on $\hat X_D$ (not $\hat Y_D$) as $\q_1=\pi/2$ in \eq{423}. For $\q_1\neq0$, \eq{430} and \eq{429} has to be rewritten as
\equ{\bra\overline{\hat\tx_D(t)\hat\tx_{B2}(t_1)}\ket=\sqrt{\frac{\e I^2}{2}}[\cos\q\cos\q_1\bra\overline{\hat Y_D(t)\hat Y_B(t_1)}\ket+\sin\q\cos\q_1\bra\overline{\hat Y_D(t)\hat X_B(t_1)}\ket\\+\sin\q_1\cos\q\bra\overline{\hat X_D(t)\hat Y_B(t_1)}\ket+\sin\q_1\sin\q\bra\overline{\hat X_D(t)\hat X_B(t_1)}\ket],}{448}
\equ{\bra\overline{\hat\tx_D(t)\hat\tx_{B1}(t_1)}\ket=\sqrt{\frac{\e I^2}{2}}[-\sin\q\cos\q_1\bra\overline{\hat Y_D(t)\hat Y_B(t_1)}\ket+\cos\q\cos\q_1\bra\overline{\hat Y_D(t)\hat X_B(t_1)}\ket\\-\sin\q\sin\q_1\bra\overline{\hat X_D(t)\hat Y_B(t_1)}\ket+\sin\q_1\cos\q\bra\overline{\hat X_D(t)\hat X_B(t_1)}\ket].}{449}
Using \eq{448} and \eq{449}, we can write
\equ{\bra\overline{\hat\tx_D(t)\hat\tx_{B1}(t_1)}\ket^2+\bra\overline{\hat\tx_D(t)\hat\tx_{B2}(t_1)}\ket^2=\{[\bra\overline{\hat Y_D(t)\hat Y_B(t_1)}\ket^2+\bra\overline{\hat Y_D(t)\hat X_B(t_1)}\ket^2]\cos^2\q_1\\+[\bra\overline{\hat X_D(t)\hat Y_B(t_1)\ket}^2+\bra\overline{\hat X_D(t)\hat X_B(t_1)\ket}^2]\sin^2\q_1\\+[\bra\overline{\hat Y_D(t)\hat Y_B(t_1)}\ket\overline{\bra\hat X_D(t)\hat Y_B(t_1)}\ket+\bra\overline{\hat Y_D(t)\hat X_B(t_1)}\ket\bra\overline{\hat X_D(t)\hat X_B(t_1)}\ket]\sin(2\q_1)\}\frac{\e}{2}I^2.}{450}
Setting $\q_1=\pi/2$ reduces \eq{450} into \eq{419} as expected. By adjusting the phase $\q=0$, we can rewrite \eq{450} as
\equ{\bra\overline{\hat\tx_D(t)\hat\tx_{B1}(t_1)}\ket^2+\bra\overline{\hat\tx_D(t)\hat\tx_{B2}(t_1)}\ket^2=\frac{\e}{2}I^2[\bra\overline{\hat Y_D(t)\hat Y_B(t_1)}\ket^2+\bra\overline{\hat Y_D(t)\hat X_B(t_1)}\ket^2].}{451}
The \eq{450} is a more general relation for any arbitrary $\q_1$. The \eq{451} and \eq{441} prove that the cross-correlation of homodyne-3 with homodyne-1 and homodyne-2 can be tuned to be a function of maximum possible correlations by tuning $\q_1$. At this point, it is impossible to say which correlations will give maximum value as it depends on the entanglement generation mechanism. For example, some entanglement schemes~\cite{schliesser-20} exhibit stronger $\bra\hat Y_D\hat Y_B\ket$ correlation in which case $\q_1=0$ (or \eq{451}) is the optimized detection scheme. Whereas, a two-mode squeezed coherent state exhibit equal correlation strength for both $\bra\hat X_D\hat X_B\ket$ and $\bra\hat Y_D\hat X_B\ket$ and hence either \eq{451} or \eq{441} will provide an optimum result.
\subsection{Quantum advantage}
The quantum advantage of this technique can be estimated by comparing it with the corresponding classical system. The corresponding classical system is built by replacing the quantum fields $\hat B$ and $\hat D$ with classical field amplitudes $\td B$ and $\td D$, respectively. The classical field amplitudes are normalized such that $|\td B|^2=\td I_B$ and $|\td D|^2=\td I_D$ with $\td I_B$ and $\td I_D$ as the classical field intensities, respectively. Adding an identical random phase fluctuation $\td\de(t)$ to both $\td B$ and $\td D$ introduces correlation in the classical system. Now the fluctuation $\td \tx_{D}$ in the output from the homodyne-3 is given as
\equ{\td\tx_{D}=\sqrt{I\td I_D}\sin\td\de(t)\approx\sqrt{I\td I_D}\td\de(t),}{458}
for $\td\de(t)<1$. Similarly, the fluctuation $\td\tx_{B1}$ ($\td\tx_{B1}$) in the output from homodyne-1 (homodyne-2) is given as
\equ{\td \tx_{B1}(t_1)=\sqrt{\frac{\e}{2}}[-\sin\q\sqrt{I\td I_B}\td\de(t_1)]+\sqrt{\frac{(1-\e)I}{2}}[\hat E^\+(t_1)e^{i\q_2}+e^{-i\q_2}\hat E(t_1)]\\+i\sqrt{\frac{I}{2}}[\hat V^\+(t_1)e^{i\q_2}-e^{-i\q_2}\hat V(t_1)],}{456}
\equ{\td \tx_{B2}(t_1)=\sqrt{\frac{\e}{2}}[\cos\q\sqrt{I\td I_B}\td \de(t_1)]+i\sqrt{\frac{(1-\e)I}{2}}[\hat E^\+(t_1)e^{i\q_2}-e^{-i\q_2}\hat E(t_1)]\\+\sqrt{\frac{I}{2}}[\hat V^\+(t_1)e^{i\q_2}+e^{-i\q_2}\hat V(t_1)].}{457}
Using \eq{458}, \eq{456}, and \eq{457}, the best performance of the classical system is given as
\equ{\bra\td \tx_D(t)\td \tx_{B1}(t_1)\ket^2+\bra\td \tx_D(t)\td \tx_{B2}(t_1)\ket^2=\frac{\e}{2}I^2\td I_D\td I_B[\bra\td \de(t)\td \de(t_1)\ket]^2.}{459}
To compare the classical result in \eq{459} with the quantum result in \eq{441}, we assume that $\hat B$ and $\hat D$ are in a two-mode squeezed coherent state, with squeeze parameter $r$. Then \eq{441} can be written as
\equ{\bra\hat \tx_D(t)\hat \tx_{B1}(t_1)\ket^2+\bra\hat \tx_D(t)\hat \tx_{B2}(t_1)\ket^2=\frac{\e}{2}I^2[\sinh r\de(t-t_1)]^2.}{460}
To make a fair comparison between classical and quantum models, in \eq{459}, we assume that $\td I_D=\td I_B=1\,\text{Hz}$ and $\bra\td \de(t)\td \de(t_1)\ket=\mathfrak{D}\de(t-t_1)$ with $\mathfrak{D}$ as correlation strength. As $\td\de(t)$ is small, the maximum value of $\mathfrak{D}$ is much less than one. In contrast, the $\sinh r$ term in \eq{460} increases exponentially with $r$, leading to quantum advantage. The smaller value of classical correlation can be compensated by increasing~\cite{helmy-23} $\td I_B$ and $\td I_D$, but then the target can easily know~\cite{helmy-19} that it is being probed. Hence, the quantum advantage is more strongly present for small $I_B$ and $I_D$, which is generally true with entangled photon generation schemes. The low intensity of the probe helps in finding the target stealthily or for imaging samples without exposing them to high optical power. Quantum mechanics ensures that the correlations in \eq{450}, \eq{451}, and \eq{441} are sharply peaked (Dirac delta function). On the other hand, classical correlations in \eq{459} must be made sufficiently sharp; otherwise, the target's distance can not be accurately estimated. Entanglement generally leads to quantum advantage as it has larger correlations~\cite{zheng-10,zubairy-06} than the classical correlations. Hence, quantum mechanics maximizes the target detection chances by creating sharper and larger correlations than its classical counterpart.
\section{Conclusion} The quantum advantage in optical ranging is studied using a cross-correlated homodyne measurement scheme. The entanglement between the probe and the idler leads to unique correlations, which filter out the probe from the environmental background. This technique is immune to environmental noise as there is no correlation between idler and the environment. The target detection and ranging efficiency are independent of the entanglement field intensity. Hence, a large intensity of entangled fields is not required to keep the quantum advantage. The target detection efficiency is proportional to the reflectivity of the target and intensity of the non-entangled reference fields in the homodyne. Hence, the low reflectivity of the target can be negated by increasing the intensity of the reference fields. All these advantages make the technique more efficient for ranging a low-reflective target immersed in a noisy environment. By heuristic arguments, we showed that the targets with reflectivity as small as $2/I^2$ can be detected and ranged.

\textbf{Acknowledgments} The author, SD, thanks Yong Li for the discussion on the measurement scheme. MJW acknowledges support from the ARC Centre of Excellence for Engineered Quantum Systems (CE170100009).\\
\textbf{Disclosures} The authors declare no conflicts of interest.\\
\textbf{Data availability} No data were generated or analyzed in the presented research.\\
\textbf{Author contribution} SD formulated the problem. SD and MJW came up with the measurement scheme to solve the problem. GG is involved with comparing the quantum illumination results with the technique described in this manuscript.
\bibliographystyle{unsrt}

\providecommand{\noopsort}[1]{}\providecommand{\singleletter}[1]{#1}%
\begin{thebibliography}{10}

\bibitem{lloyd-08}
Seth Lloyd.
\newblock Enhanced sensitivity of photodetection via quantum illumination.
\newblock {\em Science}, 321(5895):1463--1465, 2008.

\bibitem{genovesa-13}
ED~Lopaeva, I~Ruo Berchera, Ivo~Pietro Degiovanni, S~Olivares, Giorgio Brida,
  and Marco Genovese.
\newblock Experimental realization of quantum illumination.
\newblock {\em Physical review letters}, 110(15):153603, 2013.

\bibitem{gu-20}
Ranjith Nair and Mile Gu.
\newblock Fundamental limits of quantum illumination.
\newblock {\em Optica}, 7(7):771--774, 2020.

\bibitem{shapiro-08}
Si-Hui Tan, Baris~I Erkmen, Vittorio Giovannetti, Saikat Guha, Seth Lloyd,
  Lorenzo Maccone, Stefano Pirandola, and Jeffrey~H Shapiro.
\newblock Quantum illumination with gaussian states.
\newblock {\em Physical review letters}, 101(25):253601, 2008.

\bibitem{park-22}
Eylee Jung and DaeKil Park.
\newblock Quantum illumination with three-mode gaussian state.
\newblock {\em Quantum Information Processing}, 21(2):71, 2022.

\bibitem{pirandola-20}
Athena Karsa, Gaetana Spedalieri, Quntao Zhuang, and Stefano Pirandola.
\newblock Quantum illumination with a generic gaussian source.
\newblock {\em Phys. Rev. Res.}, 2:023414, Jun 2020.

\bibitem{zhao-20}
Man-Hong Yung, Fei Meng, Xiao-Ming Zhang, and Ming-Jing Zhao.
\newblock One-shot detection limits of quantum illumination with discrete
  signals.
\newblock {\em npj Quantum Information}, 6(1):75, 2020.

\bibitem{assad-21}
Mark Bradshaw, Lorc\'an~O. Conlon, Spyros Tserkis, Mile Gu, Ping~Koy Lam, and
  Syed~M. Assad.
\newblock Optimal probes for continuous-variable quantum illumination.
\newblock {\em Phys. Rev. A}, 103:062413, Jun 2021.

\bibitem{guo-14}
ShengLi Zhang, JianSheng Guo, WanSu Bao, JianHong Shi, ChenHui Jin, XuBo Zou,
  and GuangCan Guo.
\newblock Quantum illumination with photon-subtracted continuous-variable
  entanglement.
\newblock {\em Physical review A}, 89(6):062309, 2014.

\bibitem{wei-23}
De~He, X.~N. Feng, and L.~F. Wei.
\newblock Sensitive enhancement of cat state quantum illumination.
\newblock {\em Opt. Express}, 31(11):17709--17715, May 2023.

\bibitem{su-yong-lee-22}
Changsuk Noh, Changhyoup Lee, and Su-Yong Lee.
\newblock Quantum illumination with definite photon-number entangled states.
\newblock {\em JOSA B}, 39(5):1316--1322, 2022.

\bibitem{kim-21}
Su-Yong Lee, Yong~Sup Ihn, and Zaeill Kim.
\newblock Quantum illumination via quantum-enhanced sensing.
\newblock {\em Phys. Rev. A}, 103:012411, Jan 2021.

\bibitem{chandrashekar-21}
Ashwith~Varadaraj Prabhu, Baladitya Suri, and CM~Chandrashekar.
\newblock Hyperentanglement-enhanced quantum illumination.
\newblock {\em Physical Review A}, 103(5):052608, 2021.

\bibitem{jeffers-21}
Hao Yang, Wojciech Roga, Jonathan~D. Pritchard, and John Jeffers.
\newblock Gaussian state-based quantum illumination with simple photodetection.
\newblock {\em Opt. Express}, 29(6):8199--8215, Mar 2021.

\bibitem{erkmen-09}
Saikat Guha and Baris~I. Erkmen.
\newblock Gaussian-state quantum-illumination receivers for target detection.
\newblock {\em Phys. Rev. A}, 80:052310, Nov 2009.

\bibitem{shapiro-17-a}
Quntao Zhuang, Zheshen Zhang, and Jeffrey~H. Shapiro.
\newblock Optimum mixed-state discrimination for noisy entanglement-enhanced
  sensing.
\newblock {\em Phys. Rev. Lett.}, 118:040801, Jan 2017.

\bibitem{blakely-21}
Jonathan~N Blakely.
\newblock Quantum illumination with a parametrically amplified idler.
\newblock {\em Physics Letters A}, 400:127319, 2021.

\bibitem{su-yong-lee-21}
Yonggi Jo, Sangkyung Lee, Yong~Sup Ihn, Zaeill Kim, and Su-Yong Lee.
\newblock Quantum illumination receiver using double homodyne detection.
\newblock {\em Physical Review Research}, 3(1):013006, 2021.

\bibitem{kenchaf-22}
Sylvain Borderieux, Arnaud Coatanhay, and Ali Khenchaf.
\newblock Estimation of the influence of a noisy environment on the binary
  decision strategy in a quantum illumination radar.
\newblock {\em Sensors}, 22(13), 2022.

\bibitem{park-23}
MuSeong Kim, Mi-Ra Hwang, Eylee Jung, and DaeKil Park.
\newblock Is entanglement a unique resource in quantum illumination?
\newblock {\em Quantum Information Processing}, 22(2):98, 2023.

\bibitem{lloyd-09}
Jeffrey~H Shapiro and Seth Lloyd.
\newblock Quantum illumination versus coherent-state target detection.
\newblock {\em New Journal of Physics}, 11(6):063045, 2009.

\bibitem{shapiro-15}
Zheshen Zhang, Sara Mouradian, Franco~NC Wong, and Jeffrey~H Shapiro.
\newblock Entanglement-enhanced sensing in a lossy and noisy environment.
\newblock {\em Physical review letters}, 114(11):110506, 2015.

\bibitem{rajgopal-09}
A.~R. Usha~Devi and A.~K. Rajagopal.
\newblock Quantum target detection using entangled photons.
\newblock {\em Phys. Rev. A}, 79:062320, Jun 2009.

\bibitem{pirandola-15}
Shabir Barzanjeh, Saikat Guha, Christian Weedbrook, David Vitali, Jeffrey~H
  Shapiro, and Stefano Pirandola.
\newblock Microwave quantum illumination.
\newblock {\em Physical review letters}, 114(8):080503, 2015.

\bibitem{fink-20}
Shabir Barzanjeh, Stefano Pirandola, David Vitali, and Johannes~M Fink.
\newblock Microwave quantum illumination using a digital receiver.
\newblock {\em Science advances}, 6(19):eabb0451, 2020.

\bibitem{balaji-19}
David Luong, CW~Sandbo Chang, AM~Vadiraj, Anthony Damini, Christopher~M Wilson,
  and Bhashyam Balaji.
\newblock Receiver operating characteristics for a prototype quantum two-mode
  squeezing radar.
\newblock {\em IEEE Transactions on Aerospace and Electronic Systems},
  56(3):2041--2060, 2019.

\bibitem{huard-23}
R.~Assouly, R.~Dassonneville, T.~Peronnin, A.~Bienfait, and B.~Huard.
\newblock Quantum advantage in microwave quantum radar.
\newblock {\em Nature Physics}, 19(10):1418--1422, 2023.

\bibitem{wang-23}
Gewei Qian, Xingqi Xu, Shun-An Zhu, Chenran Xu, Fei Gao, V.~V. Yakovlev,
  Xu~Liu, Shi-Yao Zhu, and Da-Wei Wang.
\newblock Quantum induced coherence light detection and ranging.
\newblock {\em Phys. Rev. Lett.}, 131:033603, Jul 2023.

\bibitem{padgett-20}
Thomas Gregory, P-A Moreau, Ermes Toninelli, and Miles~J Padgett.
\newblock Imaging through noise with quantum illumination.
\newblock {\em Science advances}, 6(6):eaay2652, 2020.

\bibitem{guo-22}
Rongyu Wei, Jun Li, Weihao Wang, Songhao Meng, Baoshan Zhang, and Qinghua Guo.
\newblock Comparison of snr gain between quantum illumination radar and
  classical radar.
\newblock {\em Opt. Express}, 30(20):36167--36175, Sep 2022.

\bibitem{chandrashekar-23}
K.~Muhammed Shafi, A.~Padhye, and C.~M. Chandrashekar.
\newblock Quantum illumination using polarization-path entangled single photons
  for low reflectivity object detection in a noisy background.
\newblock {\em Opt. Express}, 31(20):32093--32104, Sep 2023.

\bibitem{helmy-22}
Phillip~S. Blakey, Han Liu, Georgios Papangelakis, Yutian Zhang, Zacharie~M.
  L{\'e}ger, Meng~Lon Iu, and Amr~S. Helmy.
\newblock Quantum and non-local effects offer over 40 db noise resilience
  advantage towards quantum lidar.
\newblock {\em Nature Communications}, 13(1):5633, 2022.

\bibitem{helmy-19}
Han Liu, Daniel Giovannini, Haoyu He, Duncan England, Benjamin~J. Sussman,
  Bhashyam Balaji, and Amr~S. Helmy.
\newblock Enhancing lidar performance metrics using continuous-wave photon-pair
  sources.
\newblock {\em Optica}, 6(10):1349--1355, Oct 2019.

\bibitem{england-18}
Bhashyam Balaji and Duncan England.
\newblock Quantum illumination: A laboratory investigation.
\newblock In {\em 2018 International Carnahan Conference on Security Technology
  (ICCST)}, pages 1--4. IEEE, 2018.

\bibitem{wilson-19}
C.~W.~Sandbo Chang, A.~M. Vadiraj, J.~Bourassa, B.~Balaji, and C.~M. Wilson.
\newblock {Quantum-enhanced noise radar}.
\newblock {\em Applied Physics Letters}, 114(11):112601, 03 2019.

\bibitem{sussman-19}
Duncan~G. England, Bhashyam Balaji, and Benjamin~J. Sussman.
\newblock Quantum-enhanced standoff detection using correlated photon pairs.
\newblock {\em Phys. Rev. A}, 99:023828, Feb 2019.

\bibitem{sussman-20}
Yingwen Zhang, Duncan England, Andrei Nomerotski, Peter Svihra, Steven
  Ferrante, Paul Hockett, and Benjamin Sussman.
\newblock Multidimensional quantum-enhanced target detection via
  spectrotemporal-correlation measurements.
\newblock {\em Phys. Rev. A}, 101:053808, May 2020.

\bibitem{ren-20}
Lorenzo Maccone and Changliang Ren.
\newblock Quantum radar.
\newblock {\em Physical Review Letters}, 124(20):200503, 2020.

\bibitem{sanz-22}
Maximilian Reichert, Roberto Di~Candia, Moe~Z Win, and Mikel Sanz.
\newblock Quantum-enhanced doppler lidar.
\newblock {\em npj Quantum Information}, 8(1):147, 2022.

\bibitem{zhuang-21}
Quntao Zhuang.
\newblock Quantum ranging with gaussian entanglement.
\newblock {\em Physical Review Letters}, 126(24):240501, 2021.

\bibitem{knott-21}
Ricardo Gallego~Torrom{\'e}.
\newblock Quantum illumination with multiple entangled photons.
\newblock {\em Advanced Quantum Technologies}, 4(11):2100101, 2021.

\bibitem{kok-21}
Zixin Huang, Cosmo Lupo, and Pieter Kok.
\newblock Quantum-limited estimation of range and velocity.
\newblock {\em PRX Quantum}, 2(3):030303, 2021.

\bibitem{shapiro-17}
Quntao Zhuang, Zheshen Zhang, and Jeffrey~H Shapiro.
\newblock Entanglement-enhanced lidars for simultaneous range and velocity
  measurements.
\newblock {\em Physical Review A}, 96(4):040304, 2017.

\bibitem{dowling-08}
Jonathan~P. Dowling.
\newblock Quantum optical metrology – the lowdown on high-n00n states.
\newblock {\em Contemporary Physics}, 49(2):125--143, 2008.

\bibitem{girvin-10}
K.~B\o{}rkje, A.~Nunnenkamp, B.~M. Zwickl, C.~Yang, J.~G.~E. Harris, and S.~M.
  Girvin.
\newblock Observability of radiation-pressure shot noise in optomechanical
  systems.
\newblock {\em Phys. Rev. A}, 82:013818, Jul 2010.

\bibitem{schliesser-20}
Junxin Chen, Massimiliano Rossi, David Mason, and Albert Schliesser.
\newblock Entanglement of propagating optical modes via a mechanical interface.
\newblock {\em Nature Communications}, 11(1):943, Feb 2020.

\bibitem{helmy-23}
Han Liu, Changhao Qin, Georgios Papangelakis, Meng~Lon Iu, and Amr~S. Helmy.
\newblock Compact all-fiber quantum-inspired lidar with over 100 db noise
  rejection and single photon sensitivity.
\newblock {\em Nature Communications}, 14(1):5344, 2023.

\bibitem{zheng-10}
Mark Hillery, Ho~Trung Dung, and Hongjun Zheng.
\newblock Conditions for entanglement in multipartite systems.
\newblock {\em Phys. Rev. A}, 81:062322, Jun 2010.

\bibitem{zubairy-06}
Mark Hillery and M.~Suhail Zubairy.
\newblock Entanglement conditions for two-mode states.
\newblock {\em Phys. Rev. Lett.}, 96:050503, Feb 2006.

\end{thebibliography}
\providecommand{\noopsort}[1]{}\providecommand{\singleletter}[1]{#1}%

\end{document}